\documentclass[%
 reprint,
 amsmath,amssymb,
 aps,
]{revtex4-1}

\usepackage{graphicx}
\usepackage{dcolumn}
\usepackage{bm}

\newcommand{\bea}{\begin{eqnarray}}
\newcommand{\eea}{\end{eqnarray}}
\newcommand{\ket}[1]{\left|{#1}\right\rangle}
\newcommand{\bra}[1]{\left\langle{#1}\right|}
\newcommand{\aver}[1]{\left\langle{#1}\right\rangle}

\usepackage{hyperref}
\usepackage{caption} 
\usepackage{booktabs,siunitx}
\usepackage{epsfig}

\usepackage{enumerate}
\usepackage{float}

\begin{document}


\title{Recurrence network analysis in a model tripartite quantum system}

\author{Pradip Laha}
 \altaffiliation[]{pradip@physics.iitm.ac.in}
\author{S Lakshmibala}%
\affiliation{%
 Department of Physics, IIT Madras, Chennai 600036, India\\}%
\author{V Balakrishnan}%
\affiliation{Department of Physics, IIT Madras, Chennai 600036, India\\}%




\date{\today}

\begin{abstract}
In a novel approach to quantum dynamics, we apply the tools of recurrence network analysis to the dynamics of the quantum mechanical expectation values of observables. We construct and analyse $\epsilon$-recurrence networks from the time-series data of the mean photon number in a model tripartite quantum system governed by a nonlinear Hamiltonian. The role played by the intensity-dependent field-atom coupling in the dynamics is investigated. Interesting features emerge as a function of a parameter characterising this intensity-dependent coupling in both the short-time and the long-time dynamics. In particular, we examine the manner in which standard measures of network theory such as the average path length, the link density  and the clustering coefficient depend on this parameter. 

\begin{description}
\item[PACS numbers] 05.45.T;  42.50.-p; 05.45.-a
\end{description}
\end{abstract}

\maketitle


\section{Introduction}
\label{sec:intro}
Atom optics provides an ideal platform for exploring the rich spectrum of nonclassical effects displayed by the state of a physical system during temporal evolution.  These effects include  wave packet revival phenomena \cite{robinett}, squeezing properties of the state of the system \cite{hong}, and  changes in the degree of entanglement between the subsystems of the total system \cite{eberly}.  These effects have been examined extensively in models of atom-field interaction in the presence of field nonlinearities  as well as  intensity-dependent couplings (IDC). For instance, the changes with time of the quadrature 
squeezing and entropic squeezing of the state of 
the system or that of a subsystem have  been  
determined  directly from relevant tomograms  \cite{sharmila1,laha3}.  Sudden death of the entanglement between subsystems \cite{eberly},  as well as  
the collapse of the 
entanglement  to a constant non-zero value over a significant time-interval \cite{laha1,laha3}, have been identified theoretically in multipartite models of radiation fields interacting with atomic media. In the examples considered,  the full system is 
an isolated quantum system in a pure state that evolves 
unitarily under a self-adjoint Hamiltonian. Each  of its  
subsystems, however,  is effectively 
an open quantum system interacting with the rest of the system. These interactions are lossy, in general. Hence the subsystem dynamics, governed by the appropriate  reduced density matrix, 
 mimics that of a dissipative system. As a consequence, the dynamics of expectation values of appropriate subsystem operators could display, in principle, a wide range of interesting properties ranging from ergodicity to 
exponential sensitivity. 
 
 In  Ref. \cite{laha1}, the dynamics of a  tripartite system comprising a $\Lambda$-atom interacting with two radiation fields with inherent nonlinearities and field-atom IDC has been  investigated. For initially unentangled field and atom states, the subsequent state 
 of the system has been shown to exhibit the full range of nonclassical effects mentioned above.  In particular, the manner in which the degree of coherence of the initial field states, the nature of the nonlinearities in the fields, and the precise form of IDC affect the occurrence of nonclassical features  have been examined.  Over the time interval when entanglement collapses to a fixed non-zero value, the mean and variance of the photon number corresponding to either of 
 the field modes also settle down to  fixed values. A novel feature that appears in this model is the occurrence of a bifurcation cascade. This crucially depends on the form of the IDC, and the choice of the initial field state which is a standard coherent state (CS) $\ket{\alpha}, \,
 \alpha \in \mathbb{C}$. In the photon number basis $\{\ket{n}\}$, 
 we have $\ket{\alpha} = e^{- |\alpha|^{2}/2} \sum_{0}^{\infty}
 \alpha^{n} \ket{n}/ (n!)^{1/2}$. If $N = a^{\dagger} a$ is  the photon number operator corresponding to one of the two fields, an IDC of the form $f(N) = (1 + \kappa\, N)^{1/2}$ leads to a  bifurcation cascade displayed by the mean photon number, which is very sensitive to the value of the intensity parameter $\kappa$ ($0\le \kappa \le 1$).  This particular form of IDC is interesting from a group-theoretic 
 point of view, as it interpolates between the Heisenberg-Weyl algebra for the field operators $a$ and $a^{\dagger}$ when  $\kappa=0$, and 
  the $SU(1, 1)$ algebra when  $\kappa = 1$ \cite{siva}. Intermediate values of $\kappa$ correspond to a deformed $SU(1, 1)$ operator algebra.  With very small changes in $\kappa$, the mean photon number displays very different temporal behaviour over the same given 
  time interval reckoned from the initial instant of time. This varies from collapse of the mean photon number to a constant non-zero value over the interval mentioned, to oscillatory behaviour over that interval.

 A natural question to ask is whether the long-time dynamics of the system is also as  sensitive to  the  value of $\kappa$, and if so,   do any interesting correlations appear in the dynamics over these distinctly different time scales?  By employing  the tools of time-series analysis of appropriate observables, the ergodic nature of quantum expectation values  displayed over a sufficiently long time interval has  been analysed in various quantum systems \cite{sudh_pla,sudh_epl,athreya}. Treating the mean photon number as a dynamical variable, its time series has been examined from a dynamical systems view-point  in these systems, and has been  shown to display a rich range of ergodic behaviour that is dependent on the initial state considered and on the nonlinearities that are present. However, an approach to the long-time dynamics of such quantum systems based on network analysis adds a new dimension to investigations of quantum dynamics.  In this paper we report on an extensive analysis based on this approach and the interesting correlations between the short and long time dynamics dictated by the IDC of the form given above.

  Over the years, network analysis has become a powerful tool in investigating the underlying structure and temporal behaviour of complex classical dynamical systems \cite{newman_book,cohen,newman,boccaletti,kurths_phys_rep}. Several methods  to convert the  time series of a particular classical dynamical variable into an equivalent  network exist in the literature\cite{zhang,lacasa,nicolis,marwan1,yang,xu,donner_epj}, each method capturing specific features of  the system that are 
  encoded in the time series. These include, among 
  others, dynamical transitions in the system and 
   topological properties of the attractor. In particular, the analysis of the recurrence of trajectories to specific cells in the classical phase space of dynamical variables,  based on $\epsilon$-recurrence networks, is ideal for unravelling complex bifurcation scenarios \cite{marwan1,donges}, distinguishing between chaotic and non-chaotic dynamics \cite{zou}, tracing unstable periodic orbits \cite{kurths}, defining alternative notions of fractal dimensions \cite{donner_epj},  and so on. Further, this method performs efficiently even with significantly shorter time series 
   ($\sim 100$ points) \cite{marwan1,zou} compared to 
   other methods. Hence, apart from analysing model systems, 
   $\epsilon$-recurrence-based network methods have found successful applications in the analysis of nonlinear time series that occur in widely different areas  such as medicine \cite{marwan2,ramirez}, paleo-climate records \cite{zou,gao1,donges}, astrophysics \cite{zolotova},  
   time -dependence of available infrastructures \cite{paolo}, and so on. In this work, we apply network analysis to the tripartite quantum system mentioned earlier. 
    An important reason for such investigations is that large data sets are available and a network analysis most often helps to capture the important physics buried in these sets by  a judicious choice of a considerably smaller set. This is also expected to facilitate machine learning programs. 

At present, even in quantum systems such large data sets need to be examined.  Due to the inherent nature of a quantum system, extraction of valuable information from quantum data sets poses a serious challenge. Several approaches are being attempted for this purpose. For instance, topological features that are manifest in a small data set  are identified and investigations on whether they persist at all scales are being carried out \cite{seth_lloyd}.  Again,  effective simplified networks have been suggested  to identify crucial features of entanglement percolation in large data sets pertaining to quantum information processing (see, for instance, \cite{acin,cuquet,perseguers}). 

In all quantum systems the outcomes of measurement are the expectation values of appropriate quantum observables, and hence these play the role of dynamical variables. In this paper we therefore take the approach that large data sets such as long time-series of quantum observables can be effectively managed through network analysis where a significantly smaller set  of values from the time-series are selected. We identify appropriate measures from network theory which capture the essence of the dynamics of the tripartite quantum system considered.

The plan of the rest of the paper is the following: We first 
summarise  the relevant features of the quantum model. We  then  comment  briefly on the short-time bifurcation cascade of the mean photon number as the intensity parameter is varied over a range of values.   This is followed by an outline of the details of the procedure used to construct the network from the time series. We then examine important properties of the network such as the average path length, the link density and  the clustering coefficient  in order  to assess how the long-time behaviour of the mean photon number depends on $\kappa$.  The main new feature that emerges is the strong correlation 
of this  dependence  with the $\kappa$-dependence revealed in the bifurcation cascade.  

\section{\label{sec:model} The quantum model}

\begin{figure}[h]
\centering
\includegraphics[height=3.5cm, width=6cm]{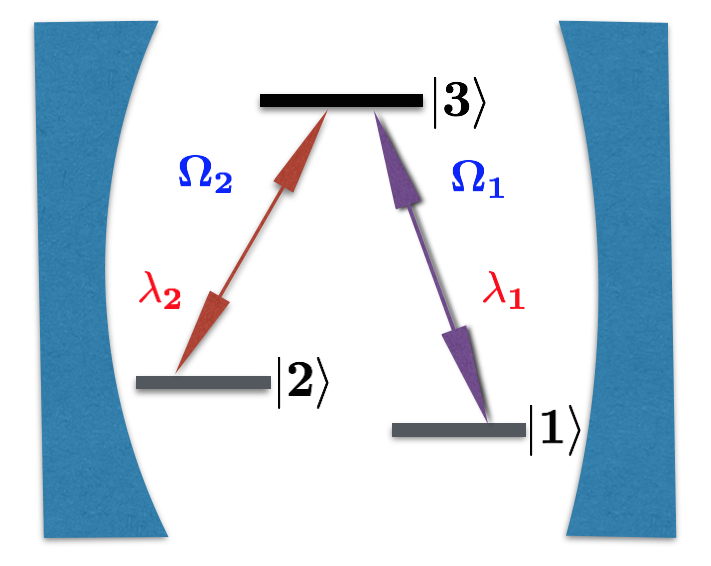}
\vspace{0ex}
\caption{Schematic diagram of the $\Lambda$ system.}
\label{lambda_atom}
\end{figure}

The tripartite quantum system considered here comprises a 
$\Lambda$-atom in a cavity, 
interacting with a probe field $F_{1}$ and a coupling field $F_{2}$ with frequencies $\Omega_{1}$ and $\Omega_{2}$ respectively. The  field annihilation and creation operators corresponding to $F_{i}$  $(i = 1, \,2)$ are $a_{i}$ and  $a^{\dagger}_{i}$. The three atomic energy eigenstates are denoted by $\ket{j}\, (j = 1,\, 2,\, 3)$ (see 
Fig.  \ref{lambda_atom}).
 \begin{figure*}
 \centering
 \includegraphics[height=6.9cm, width=4cm,angle=-90]{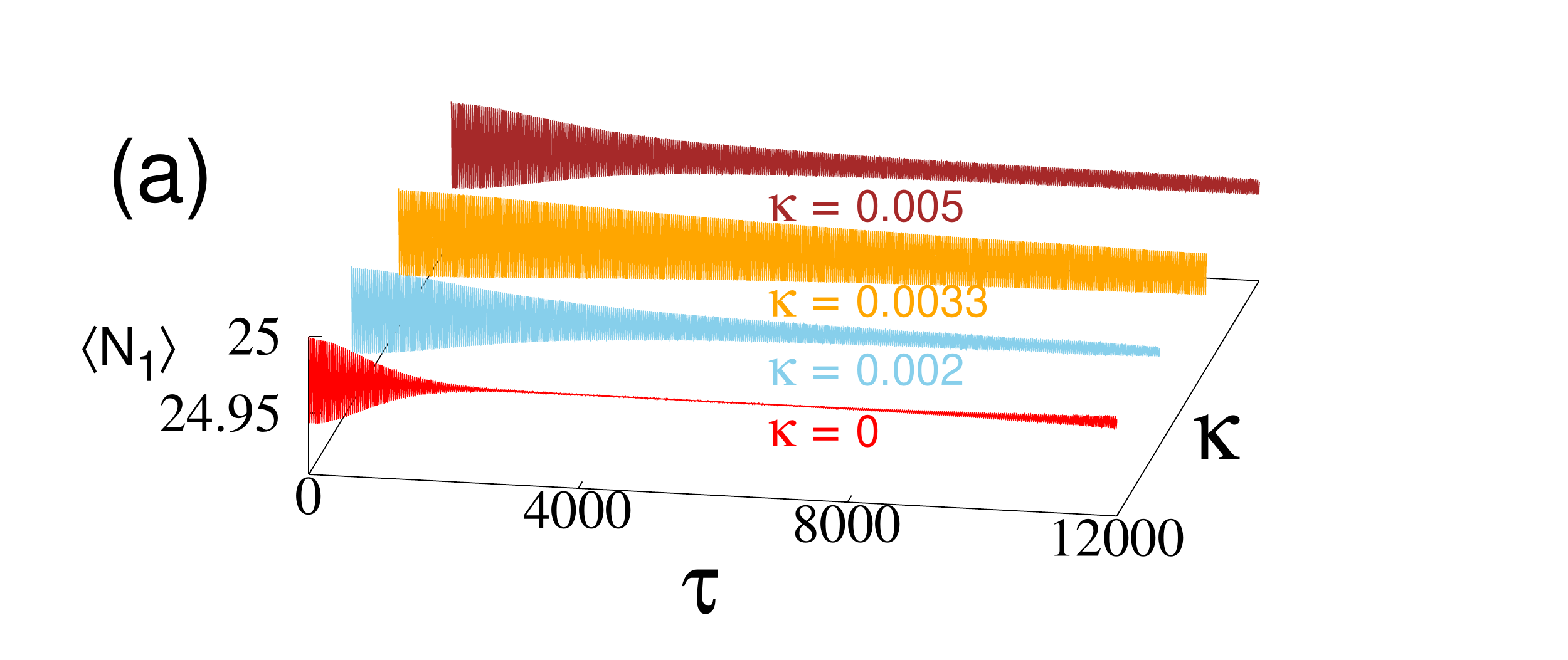}\hspace{0ex}\hspace{-11ex}
 \includegraphics[height=6.9cm, width=4cm,angle=-90]{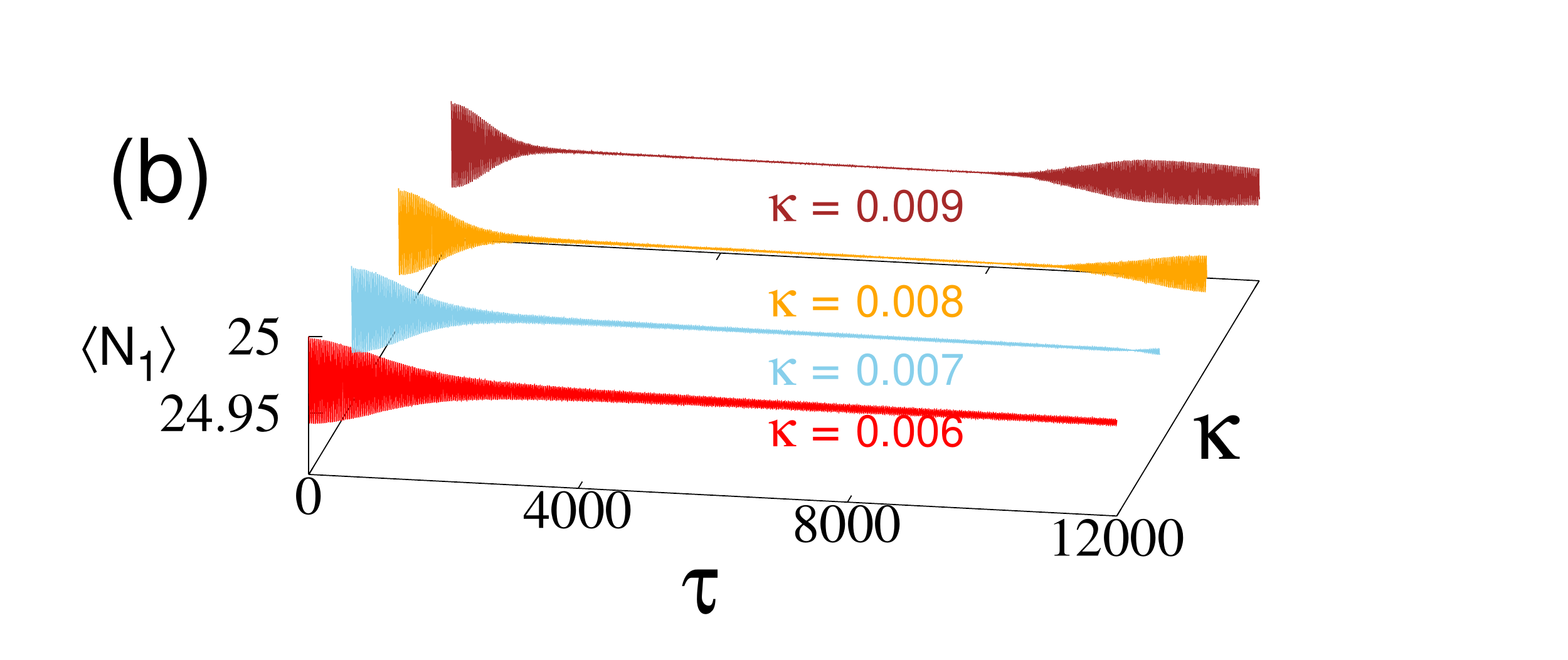}\hspace{0ex}\hspace{-11ex}
 \includegraphics[height=6.9cm, width=4cm,angle=-90]{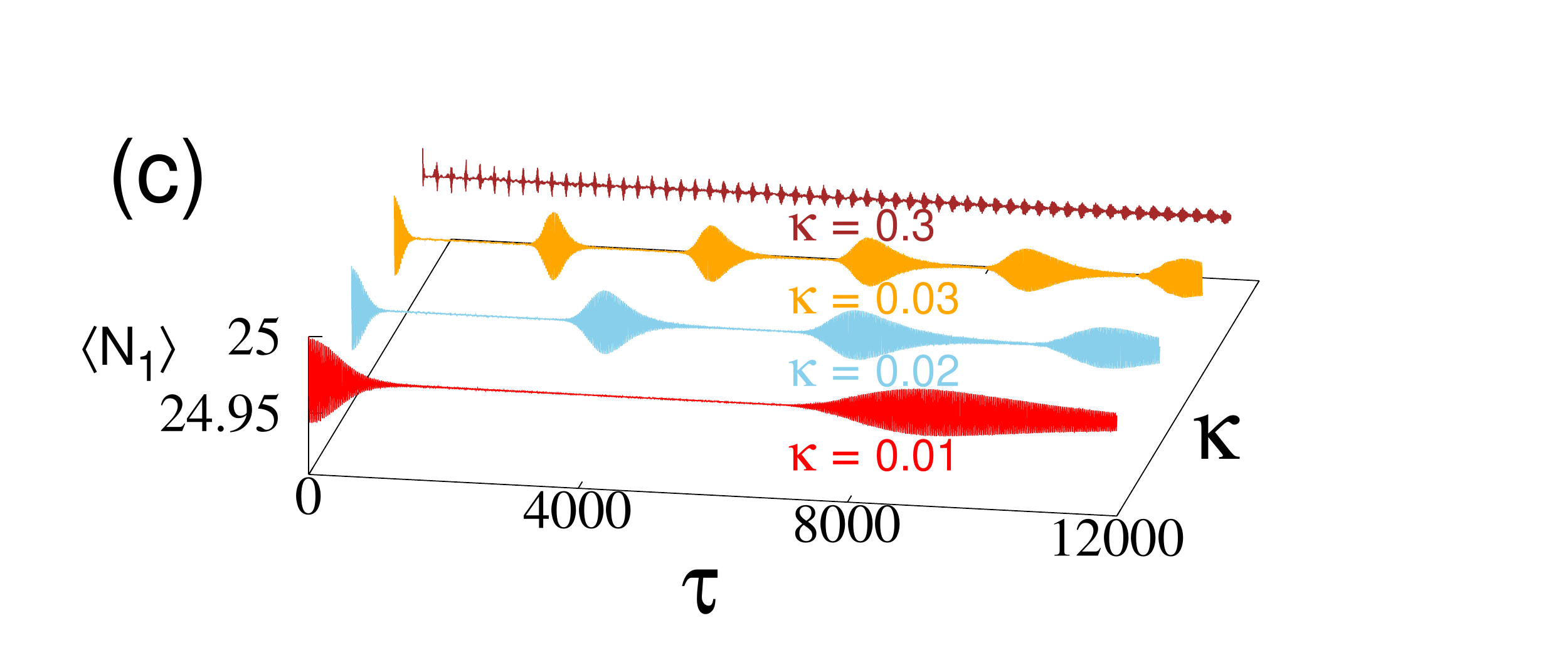}\hspace{0ex}
 \vspace{-2ex}
 \caption{$\aver{N_{1}}$ versus  $\tau$ for different values of $\kappa$. 
  Initial state $\ket{1;\alpha; \alpha}, \,\vert\alpha\vert^{2} = 25, \,\chi/\lambda = 5$.}
 \label{fig:n1_aver}
\end{figure*}

$F_{1}$ and $F_{2}$ induce the $\ket{1}\leftrightarrow \ket{3}$ and $\ket{2}\leftrightarrow \ket{3}$  transitions respectively, while the  transition $\ket{1}\leftrightarrow \ket{2}$ is dipole-forbidden. 
 The  Hamiltonian incorporating field nonlinearities and atom-field intensity-dependent couplings is (setting $\hbar = 1$) 
\begin{align}
 H  =  \sum_{j=1}^{3} \omega_{j} \sigma_{jj}  &+ 
 \sum_{i=1}^{2}\Big\{ \Omega_{i} \,a_{i}^{\dagger} a_{i}  +  \chi_{i} \,a_{i}^{\dagger 2} a_{i}^{2} \nonumber \\
                                        &+  \lambda_{i} \big[a_{i} \, f(N_{i}) \,\sigma_{3i} +  f(N_{i})\, a^{\dagger}_{i} \,\sigma_{i3})\big] \Big\}.
\label{eqn:lambda_hamiltonian}
\end{align}
 $\sigma_{jk} = \ket{j}\bra{k}$ are the Pauli spin operators, and 
 $\{\omega_{j}\}$ are positive constants.  $\chi_{i}$ 
 represents  the strength of the nonlinearity in $F_{i}$, 
  and $\lambda_{i}$ is  the atom-field coupling parameter 
  corresponding to the $\ket{3}\leftrightarrow \ket{i}$ transition 
  ($i=1, \,2$). The IDC $f(N_{i}) = (1+\kappa_{i}\, N_{i})^{1/2}$, where $N_{i} = a^{\dagger}_{i} a_{i}$.

 We denote by $\{\ket{n}\}$ and $\{\ket{m}\}$ ($n,\, m = 0,\, 1,\, 2,\, \cdots$) the photon number bases corresponding 
 respectively to the fields $F_{1}$ and $F_{2}$. The fields are initially taken to be in the CS $\ket{\alpha}$, and the atom in 
 the state  $\ket{1}$. The initial state $\ket{\psi(0)}$ of the full system  is therefore a superposition of product states 
 $\ket{1}\otimes\ket{n}\otimes \ket{m}$. The state $\ket{\psi(t)}$ at 
 any time $t > 0$ is obtained by unitary evolution of $\ket{\psi(0)}$ 
 governed by  the  Hamiltonian $H$.  The space of the 
 parameters in $H$ is quite rich. For simplicity, we  shall set 
$\lambda_{1} = \lambda_{2} = \lambda$ and 
$\chi_{1} = \chi_{2} = \chi$.   
Further, we work with zero detuning, i.e., we set 
$\omega_{3} - \omega_{i} - \Omega_{i} = 0, \, (i=1,\, 2)$,  
and  $\kappa_{2} = 0$. In the rest of this paper, we drop the suffix and denote $\kappa_{1}$ by $\kappa$.

\section{Short-time behaviour  of the mean photon number}
\label{sec:bifurcation} 

We have investigated   the temporal evolution of the system   in the strong nonlinearity regime, setting $\chi/\lambda = 5$, 
for different values of $|\alpha|^{2}$ (the initial value of the 
mean photon number). We first consider $|\alpha|^{2} = 25$.
   As  reported earlier \cite{laha1},  the mean photon number exhibits an  interesting bifurcation cascade as $\kappa$ is varied from 0 to 1 (Fig. \ref{fig:n1_aver}).  The entanglement collapse from approximately 3000 to 9000 units of scaled time $\tau$ ($= \lambda t$) for $\kappa = 0$ is replaced by a `pinched' effect over that same interval for $ \kappa = 0.002$. In contrast, for $\kappa = 0.0033$ there is a significantly larger spread in the range of values of the mean photon number, and the pinch seen for lower values of 
   $\kappa$ is absent. The qualitative behaviour of the mean photon number 
   for $\kappa = 0.005$  is very similar to that which arises for $ \kappa = 0.002$. Thus $\kappa = 0.0033$ is a special value (for the given values of 
   $\chi/\lambda$ and $\vert\alpha\vert^{2}$).  
   With  further increase in $\kappa$,  an oscillatory pattern in the mean photon number takes over, the spacing between successive crests and troughs diminishing with increasing  $\kappa$. This feature persists up to 
   $\kappa = 1$. 
 
 The  range $0< \tau \lesssim 10000$ suffices to capture all these dynamical features (including the bifurcation cascade in the dynamics of $\aver{N_{1}}$). Hence we regard this range as the {\it short-time} regime in this model. 
 
As far as the long-time dynamics is concerned, we discard the interval $0< \tau \lesssim 10000$ and  consider the range $10000\lesssim \tau \lesssim 35000$, well past the interval of the cascade. We have verified that the upper bound of $\sim 35000$ suffices to capture the dynamical features (such as the maximal Lyapunov exponent) deduced by time-series analysis going up to $\tau = 300000$. The interval $10000< \tau \lesssim 35000$ is therefore regarded as a {\it long-time} regime as far as network analysis is concerned.

In the sections that follow we establish that the long-time dynamics of the system is also very sensitive to small changes in $\kappa$, 
 and that once again $\kappa = 0.0033$ is a special value.  
 Such a correspondence between the short and long time dynamics 
 exists for all sufficiently large  values of $|\alpha|^{2}$.  
  The investigation of the long-time dynamics is based on converting the long time series data of the mean photon number into a network. 
  \begin{figure*}
 \centering
 \includegraphics[scale=0.52,angle=-0]{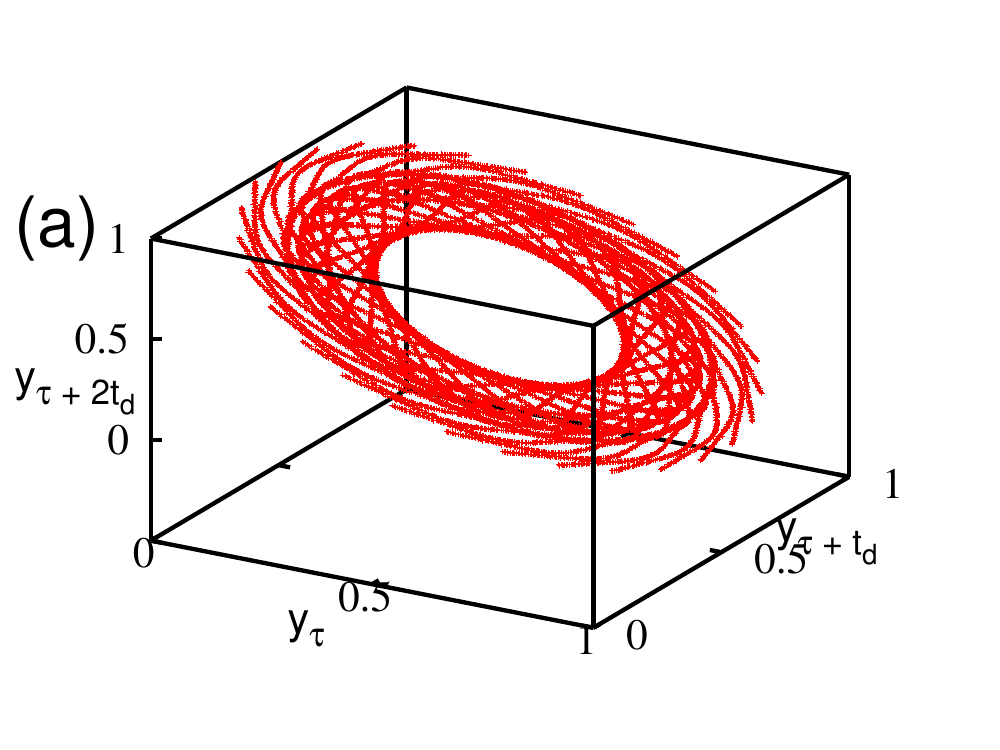}\vspace{-1ex}
 \includegraphics[scale=0.52,angle=-0]{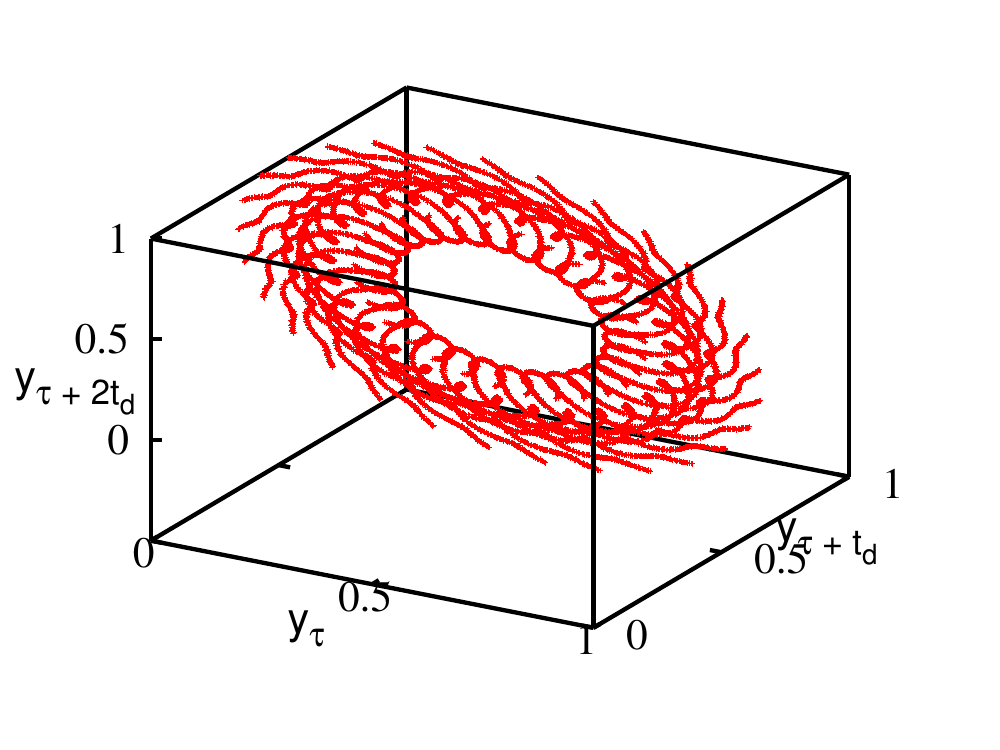}\vspace{-1ex}
 \includegraphics[scale=0.52,angle=-0]{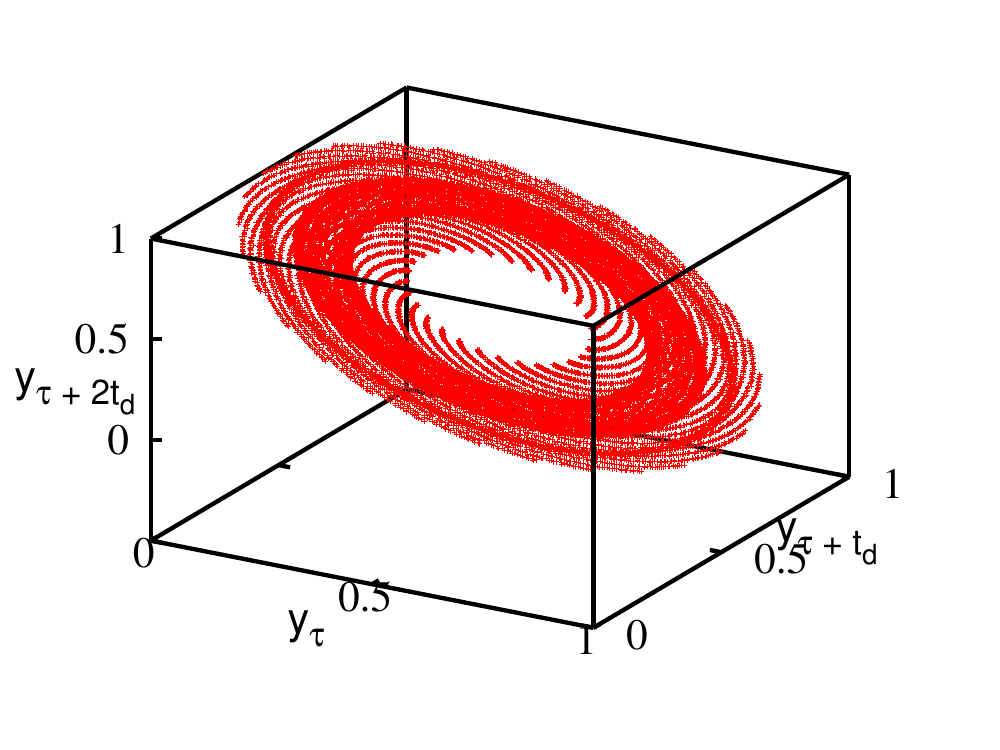}\vspace{-1ex}
 \includegraphics[scale=0.52,angle=-0]{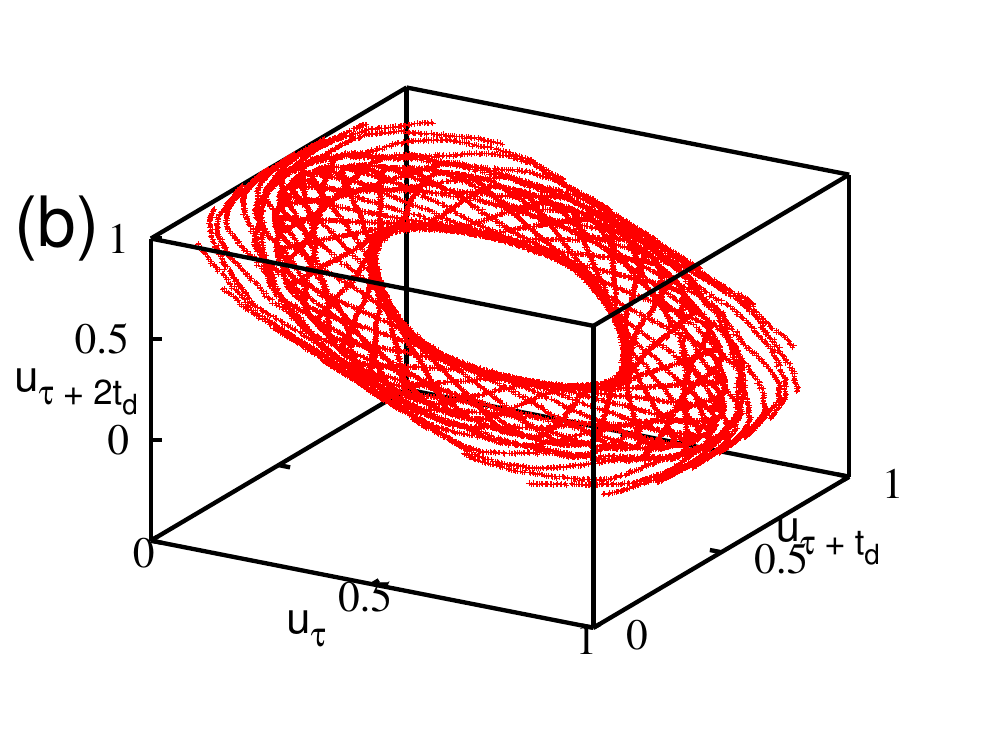}
 \includegraphics[scale=0.52,angle=-0]{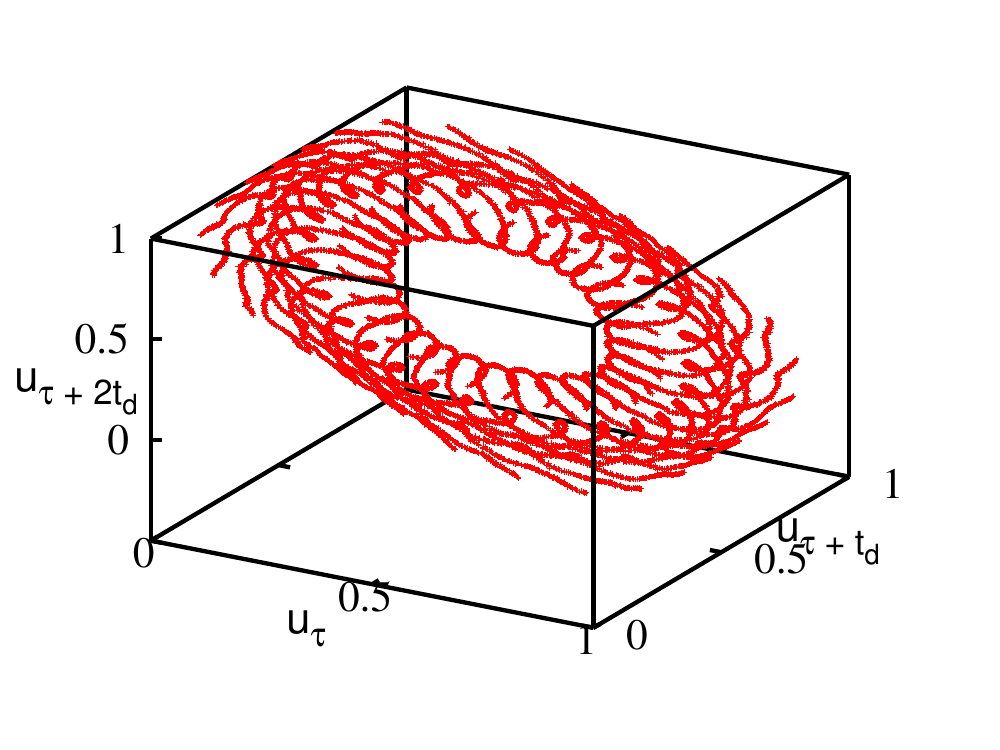}
 \includegraphics[scale=0.52,angle=-0]{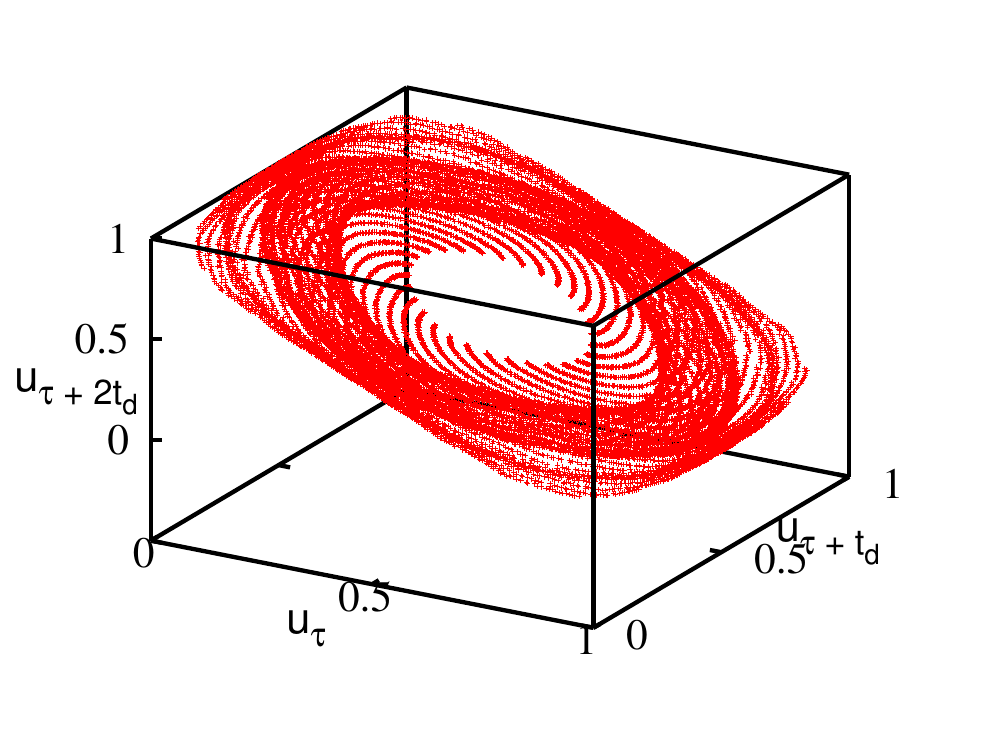}
 \vspace{-3ex}
 \caption{ Left to right: Attractors  for $\kappa = 0.0032, \,0.0033$ and 
 $0.0034$, respectively. (a) Top: rescaled time series $\{y(i)\}$. 
 (b) Bottom: uniform-deviate time series $\{u(i)\}$. 
 ($\tau = $ dimensionless time,  \,$t_{d} = $ time delay.)}
 \label{fig:att_kappa}
\end{figure*}

\section{Recurrence networks}
\label{sec:network} 
We begin with a brief description of network construction  from a time series\cite{newman_book}, in the context of the specific system of interest here. As mentioned earlier, for each value of $\kappa$, we obtain a long time series $\{s(i)\}$ ($i = 1,\, 2,\, \cdots,\, N$, where $N = 25000$)  in the interval $10000 \le \tau < 35000$. The range of variation of $\{s(i)\}$ naturally depends  on the value of $\kappa$.  In order to  facilitate comparison between the  various time series $\{s(i)\}$ obtained for different values of  $\kappa$,  each of them is  rescaled to fit into the unit interval $[0,\, 1]$.  Now  consider any one of the time series, denoted by $\{y(i)\}$.

The quantum mechanical subsystem we consider is effectively 
dissipative. Observed over a sufficiently long time, the dynamics 
appears to settle down in a very localised  region of the relevant `phase space', that we loosely term an attractor. It has been demonstrated\cite{ambika}  
in the network analysis of 
classical systems that the 
conversion of $\{y(i)\}$ into a uniform deviate time series $\{u(i)\}$ stretches the attractor in all directions  to fill the unit interval optimally, without affecting the dynamical invariants, and provides better convergence of data points. 
This procedure is considered to be helpful 
from a computational viewpoint, and so we  adopt it. 
For each $y(i)$, if $n(i)$ is the number of data points $\leq y(i)$,  the uniform deviate time-series  is given by $u(i)  = n(i)/N$. 

 Next,  employing the strategy described  for carrying out the time-series analysis for  networks \cite{kurths_phys_rep}, 
 a suitable time delay $t_{d}$ is identified.  This is the first minimum of the time-delayed mutual information.  An effective phase space of dimensions $d_{\textrm{emb}} (\ll N)$ is  
 reconstructed from $\{u(i)\}$ and $t_{d}$.
 In this phase space we then have a total of $N' = N - (d_{\textrm{emb}} - 1)t_{d}$ state vectors ${\bf x}_{j} \,\, (1 \leq j \leq N')$ given by
\begin{equation}
 {\mathbf x}_{j} = [u(j),\, u(j+t_{d}), \cdots,\, u(j + (d_{\textrm{emb}}-1)t_{d})].
 \label{eqn:delay_vec}
\end{equation}
 The dynamics takes one state vector to another and phase trajectories arise. The phase space is divided into cells of a convenient size $\epsilon$ (see 
 below).  The  recurrence properties of the phase trajectories are quantified 
 with the help of the  
 $(N'\times N')$ 
 recurrence matrix $R$  with elements   
\begin{equation}
 R_{ij} = \Theta(\epsilon - \parallel \mathbf{x_{i}}-\mathbf{x_{j}}
 \parallel), 
 \label{eqn:rij}
\end{equation}
where $\Theta$ denotes the unit step function and 
$\parallel \cdot\cdot\parallel$ is the 
Euclidean norm. $R$ is a real symmetric matrix with each  
diagonal element  equal to unity. The adjacency matrix  
is defined as $A = R-I$, where $I$ is the $(N'\times N')$ unit matrix. 
 Two state vectors (nodes) ${\mathbf x}_{i}$ and ${\mathbf x}_{j}$ ($i\ne j$) are connected iff $A_{ij} = 1$.  The network is therefore comprised of links between such connected nodes. 

Since the recurrence matrix depends on $\epsilon$, 
the network and its properties are sensitive to the value of $\epsilon$. 
It is therefore necessary to choose $\epsilon$ appropriately. 
 Too small a value of $\epsilon$ makes the network sparsely connected, 
 with an adjacency matrix that has too many vanishing off-diagonal elements.  
 On the other hand, too large value a of  $\epsilon$ would imply that 
 all  the non-diagonal elements of $A$  
 are unity,  and  the small-scale properties of the system cannot 
 then be captured. An optimal choice of the value of $\epsilon$ 
 ($= \epsilon_{c}$, say) must therefore be made. 
 Numerous methods have been suggested in the literature for obtaining $\epsilon_{c}$  \cite{gao,ambika,donner,eroglu}. Here we take the 
 approach\cite{eroglu}  summarised below.  
 
Consider the  Laplacian matrix $L = D - A$, where 
$D$ is a diagonal matrix with elements $D_{ij} = \delta_{ij}\, k_{i}$ 
(no summation over $i$) and $k_{i} = \sum_{j=1}^{N'} A_{ij}$, 
  the degree of the  node $i$. $L$ is a real symmetric matrix, 
  and each of its row sums vanishes. Together with the 
  Gershgorin circle  theorem (see, e.g., \cite{vbala_book}), these properties imply that the eigenvalues 
  of $L$ are real, non-negative, and  that at least one of the 
  eigenvalues is zero.  The recurrence network is fully connected if 
  there is a single zero eigenvalue, and the second-smallest 
  eigenvalue $l_{2}$ is positive.  By calculating $l_{2}$ for  different values 
  of $\epsilon$, the smallest value of $\epsilon$ for which $l_{2}> 0$ 
  is ascertained. We denote this value by  $\epsilon_{c}$. 
 
  In Fig. \ref{fig:att_kappa} we illustrate the manner in which the conversion of the merely rescaled time-series $\{y(i)\}$ into a uniform deviate time-series $\{u(i)\}$  stretches the attractor. Consider the case when $|\alpha|^{2} = 25$. We know 
  already from the short-term dynamics that  $\kappa =  0.0033$ is  a special value 
  in this case.   The figure displays the attractor plots for $\kappa = 0.0032$, $0.0033$, and $0.0034$ respectively.  It is evident that in the long-time 
  dynamics, too,  the plots for $\kappa = 0.0033$ are quite 
  distinct from those for even very slightly different values of $\kappa$.

In Table \ref{tab.lbl}, we list  the numerical values of $t_{d}$, $d_{\text{emb}}$ and $\epsilon_{c}$ for different values of $\kappa$, calculated from $\{u(i)\}$. We see that as $\kappa$ increases,  $t_{d}$ decreases while $d_{\text{emb}}$ increases gradually. We note also that the value of $\epsilon_{c}$ is minimum for $\kappa = 0.0033$.

\begin{table}
\caption{$t_{d}$, $d_{\text{emb}}$ and $\epsilon_{c}$ for different values of $\kappa$.}
\label{tab.lbl}
\begin{center}
\begin{tabular}{lccr}
\noalign{\smallskip} \hline \hline \noalign{\smallskip}
 $\mathbf{\kappa}$  & $\mathbf{t_{d}}$  & $\mathbf{d_{\text{emb}}}$  & $\mathbf{\epsilon_{c}}$\\
\hline \vspace{0.01ex}
    0		                  & 8		            & 3                    & 0.025\\
    0.0012   		 & 7	                    & 3	            & 0.030 \\
    0.0032   		 & 7	                    & 3	            & 0.025 \\
    0.0033   		 & 7	                    & 3	            & 0.020 \\
    0.0034   		 & 7	                    & 3	            & 0.025 \\
    0.07	                  & 4		            & 3		    & 0.060\\
    0.1        		 & 3	                    & 6	            & 0.200 \\
\noalign{\smallskip} \hline \noalign{\smallskip}
\end{tabular}
\end{center}
\end{table}


\begin{figure*}
 \centering
 \includegraphics[height=3.2cm, width=4.6cm]{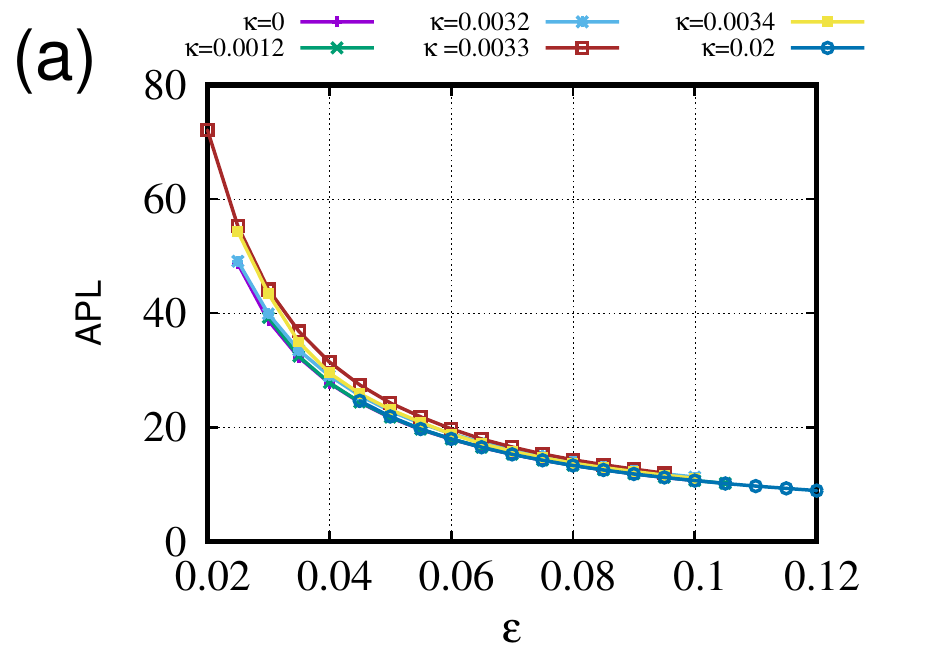}\hspace{-3ex}
 \includegraphics[height=3.2cm, width=4.6cm]{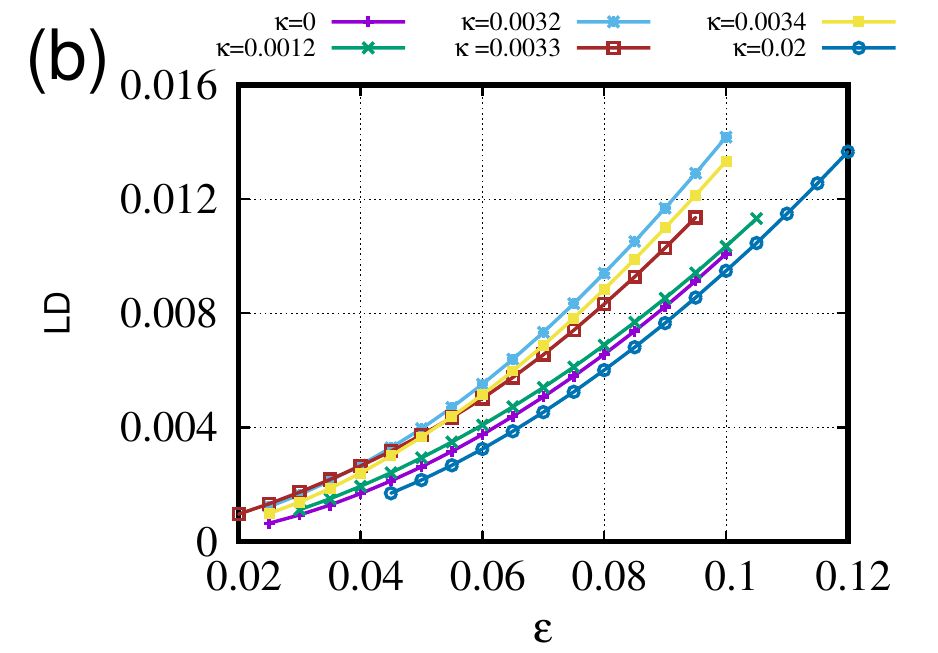}\hspace{-2ex}
 \includegraphics[height=3.2cm, width=4.6cm]{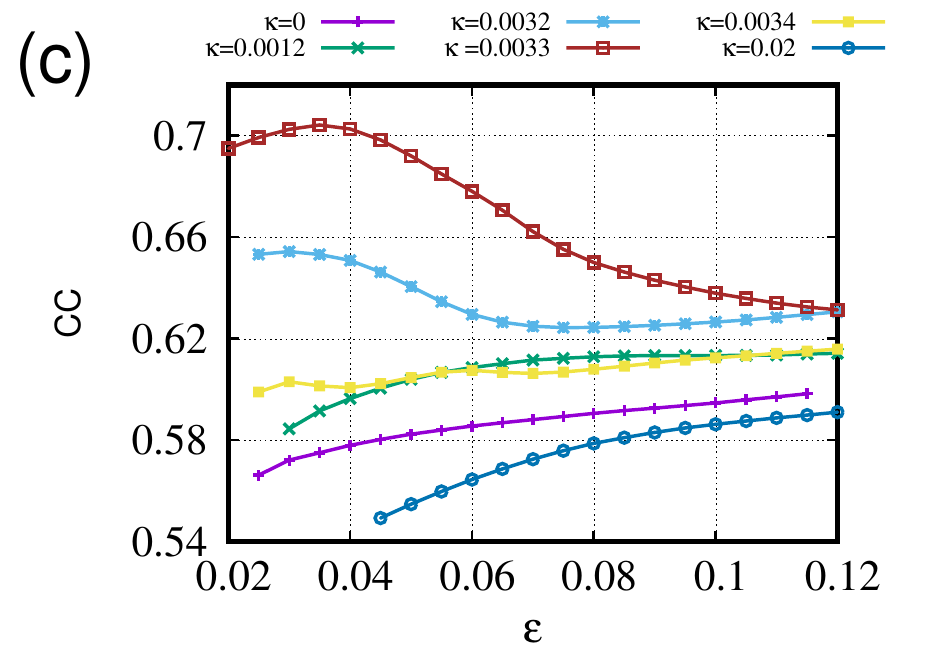}\hspace{-2ex}
 \includegraphics[height=3.2cm, width=4.6cm]{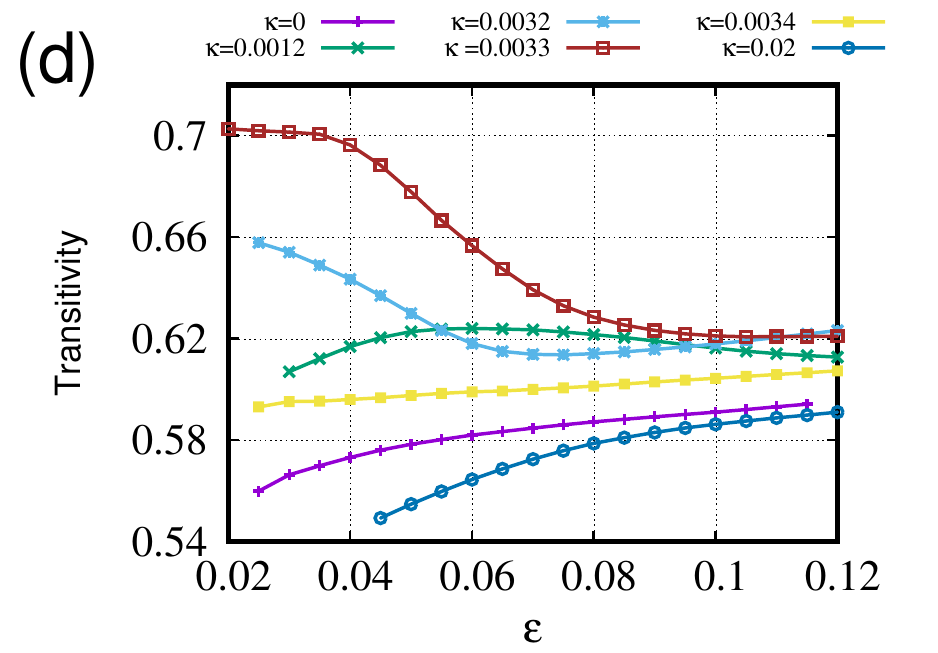}
 \vspace{-1ex}
 \caption{(a)  APL, (b) LD, (c)  CC and (d) Transitivity versus  $\epsilon$ for different values of 
 $\kappa$.}
 \label{fig:apl_den_cc_eps}
\end{figure*}

\section{Network analysis}
\label{net_res}

We now proceed to investigate the details of the networks obtained 
in the foregoing 
for different values of $\kappa$. 
We have estimated the average path lengths, link densities, clustering coefficients, transitivities, degree distributions and assortativities  which characterise  the dynamics underlying  generic networks.  (For ready reference we state  
the definitions of  some of these  quantities\cite{newman_book,boccaletti,kurths_phys_rep,strogatz}  which are relevant to our present purposes.)
The sensitivity  of the network to the value of $\kappa$ is also examined.

The average path length APL (or the characteristic path length) of a network of $P$ nodes is given by
\begin{equation}
 \text{APL} = 1/[P(P-1)] \sum_{i,j}^{P} d_{ij},
 \label{eqn:apl}
\end{equation}
where $d_{ij}$ is the {\em  shortest} path length connecting nodes $i$ and $j$.
 The link density LD of the network is defined as 
\begin{equation}
 \text{LD} = 1/[P(P-1)] \sum_{i}^{P} k_{i},
 \label{eqn:apl}
\end{equation}
where $k_{i}$ is the degree of the node $i$ (defined earlier). 
The local transitivity characteristics of a complex network are quantified by the local clustering coefficient, which measures the probability that two randomly chosen neighbours of a given node $i$ are directly connected. For finite networks, this probability is given by 
\begin{equation}
 C_{i} = 1/[k_{i}(k_{i}-1)]  \sum_{j,k}^{P} A_{jk} \, A_{ij}\, A_{ik}.
 \label{eqn:lcc}
\end{equation}
The global clustering coefficient (CC) is defined as the arithmetic mean of the local clustering coefficients taken over all the nodes of the network, i.e.,
\begin{equation}
 \text{CC} =  (1/P)  \sum_{i}^{P} C_{i}.
 \label{eqn:cc}
\end{equation}
The transitivity $\mathcal{T}$ of a network is defined as 
\begin{equation}
 \mathcal{T} =   \frac{\sum_{i,j,k}^{P}   A_{ij}\,A_{jk}\, A_{ki}}{\sum_{i,j,k}^{P} A_{ij} \, A_{ki}}.
 \label{eqn:transitivity}
\end{equation}

The dependence of  APL, LD,  CC and transitivity on $\epsilon $ (where $\epsilon \ge \epsilon_{c}$) for $|\alpha|^{2} = 25$ is shown in Fig. \ref{fig:apl_den_cc_eps}. For all values of $\kappa$, APL decreases (Fig.  \ref{fig:apl_den_cc_eps}(a)) and  LD increases (Fig.  \ref{fig:apl_den_cc_eps}(b)) with increasing $\epsilon$. This is to be expected: the number of links in the networks increases with increasing  $\epsilon$,  and this results in shorter average path length, and larger link densities.  However decrease in CC  for sufficiently large $\epsilon$  indicates  that the  closed loops in the network  do not increase significantly. In the reconstructed dynamics this would suggest that periodic orbits are few, and ergodic and/or long-period trajectories are more prevalent.   An interesting feature is that both CC and transitivity are very sensitive to the value of $\kappa$.  For  values  close to  $0.0033$ the manner in which these change with $\epsilon$ is distinctly different from that for other values of $\kappa$  (Figs.  \ref{fig:apl_den_cc_eps}(c)and (d)). Further, both these measures pass through a maximum for $\kappa = 0.0033$, attaining a  value that is significantly larger than that  in all other cases.  Thus we have identified that CC and transitivity capture the uniqueness of the value $\kappa = 0.0033$.  We have verified that  like APL and LD the degree distribution and assortativity  are also not good indicators. For completeness, we have reported the degree distribution for $|\alpha|^{2} = 25$ and for different values of $\kappa$ (Figs. \ref{fig:deg_dist}(a)-(d)).  With changes in the values of  $|\alpha|^{2}$ these conclusions do not change.

\begin{figure*}
 \centering
 \includegraphics[height=7cm, width=4cm,angle=-90]{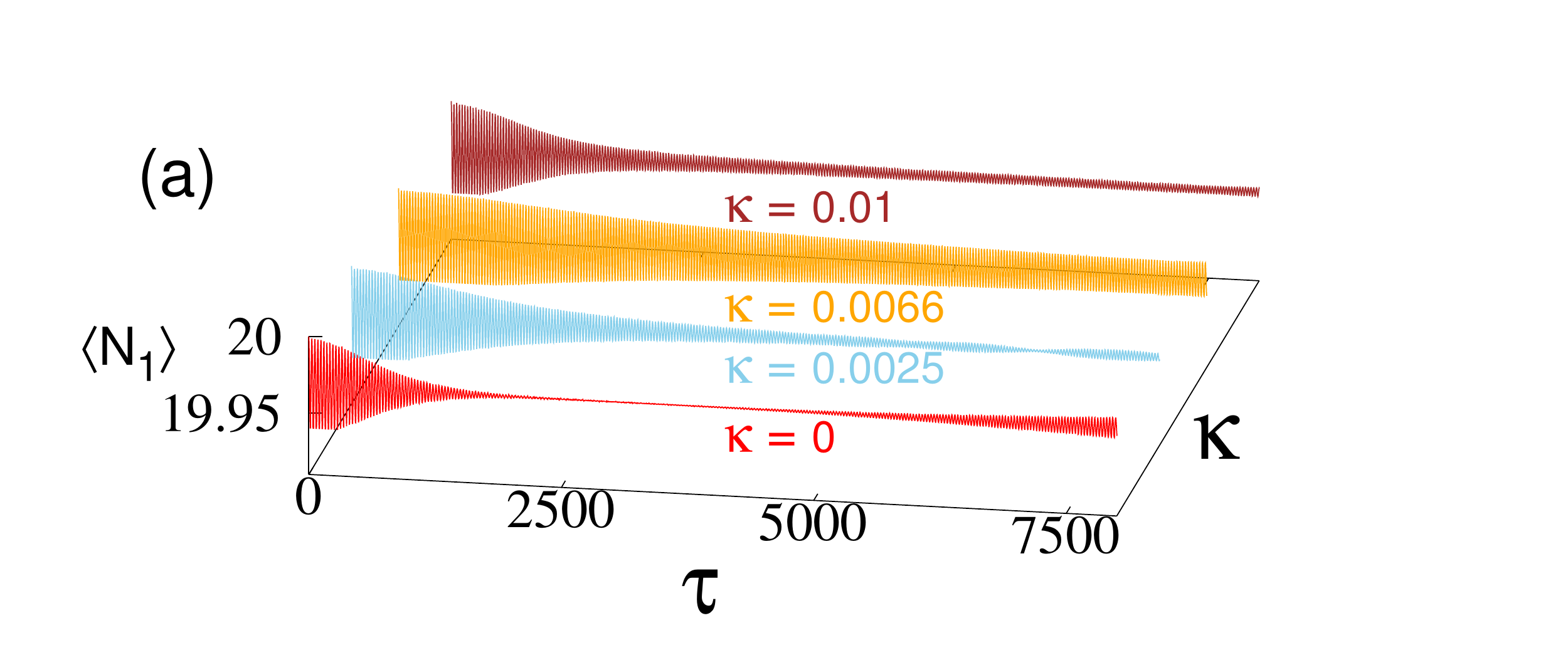}\hspace{-12ex}
 \includegraphics[height=7cm, width=4cm,angle=-90]{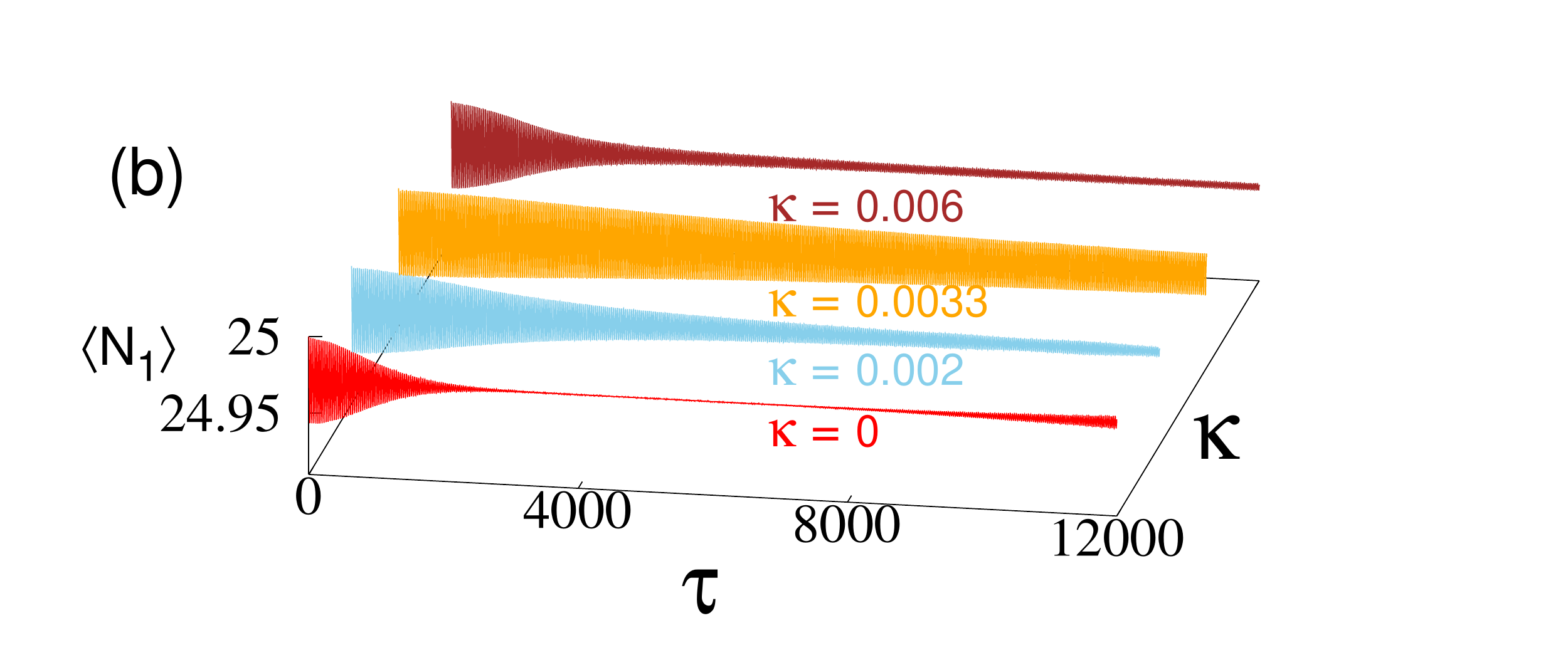}\hspace{-12ex}
 \includegraphics[height=7cm, width=4cm,angle=-90]{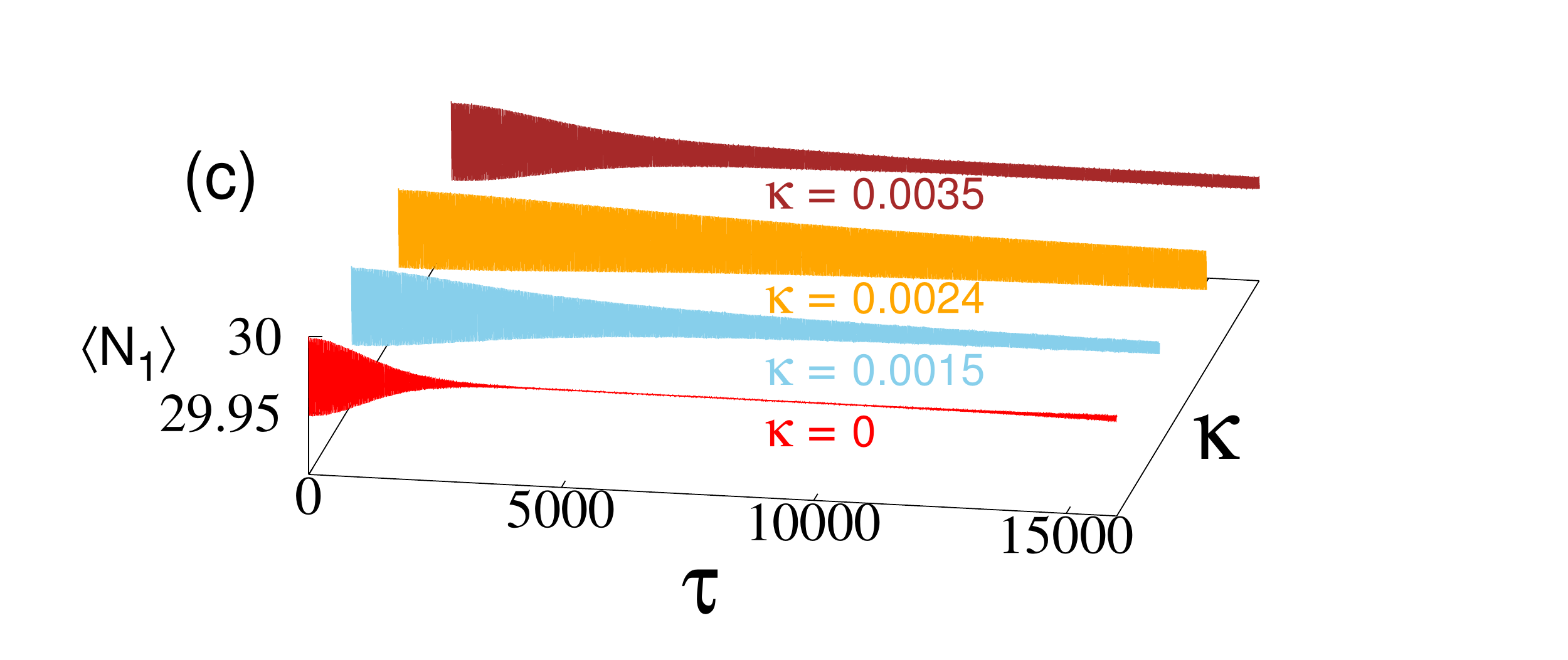}\vspace{-0ex}
 \includegraphics[height=3cm, width=5.5cm]{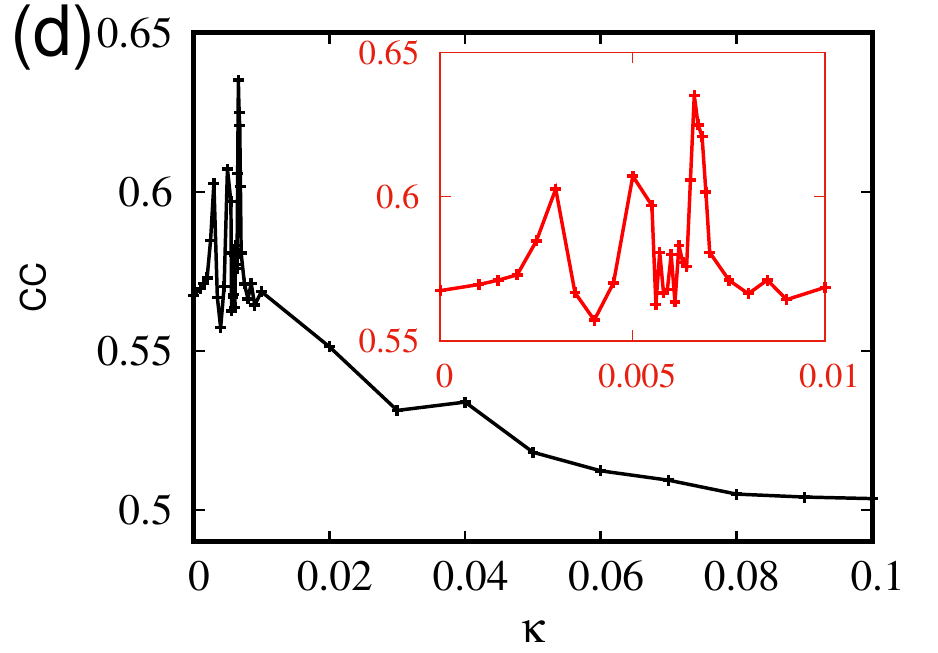}
 \includegraphics[height=3cm, width=5.5cm]{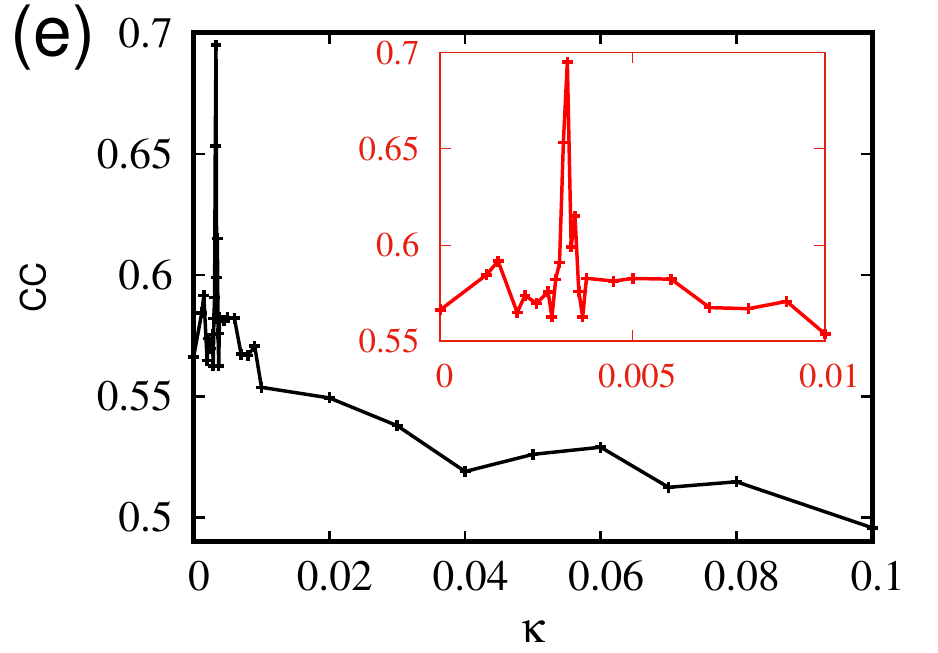}
 \includegraphics[height=3cm, width=5.5cm]{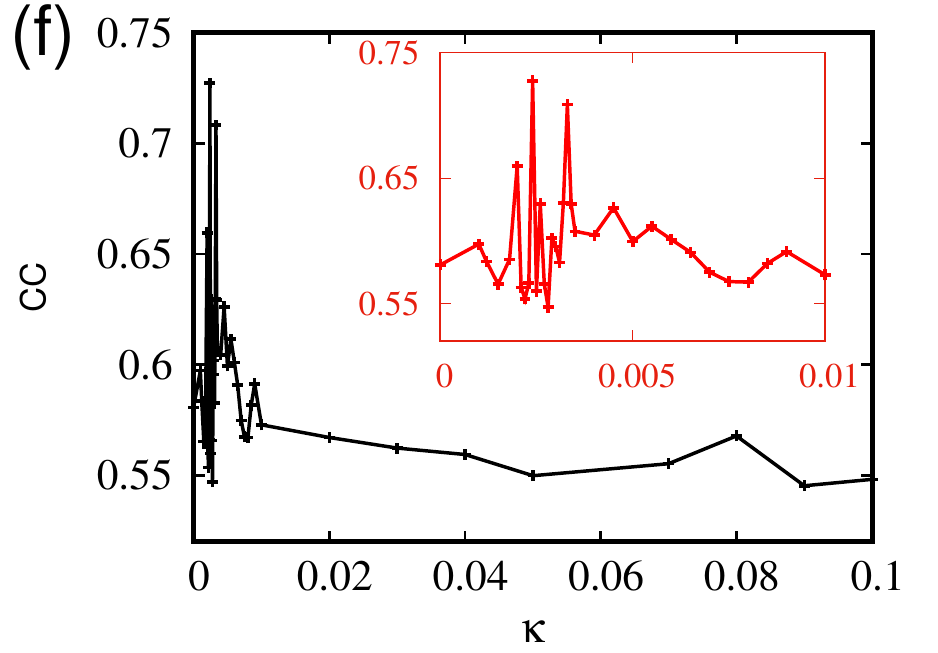}
 \includegraphics[height=3cm, width=5.5cm]{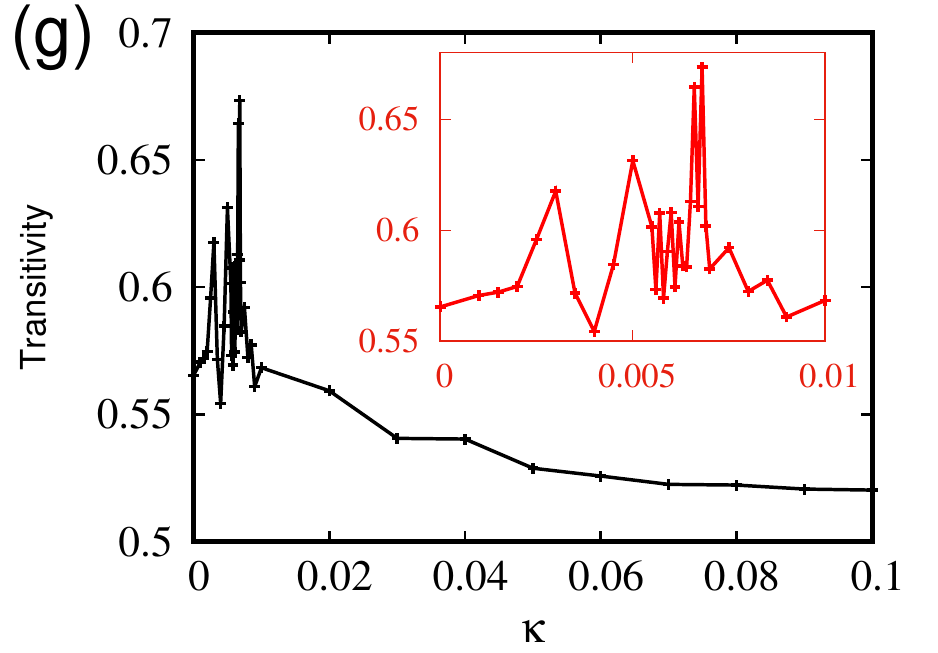}
 \includegraphics[height=3cm, width=5.5cm]{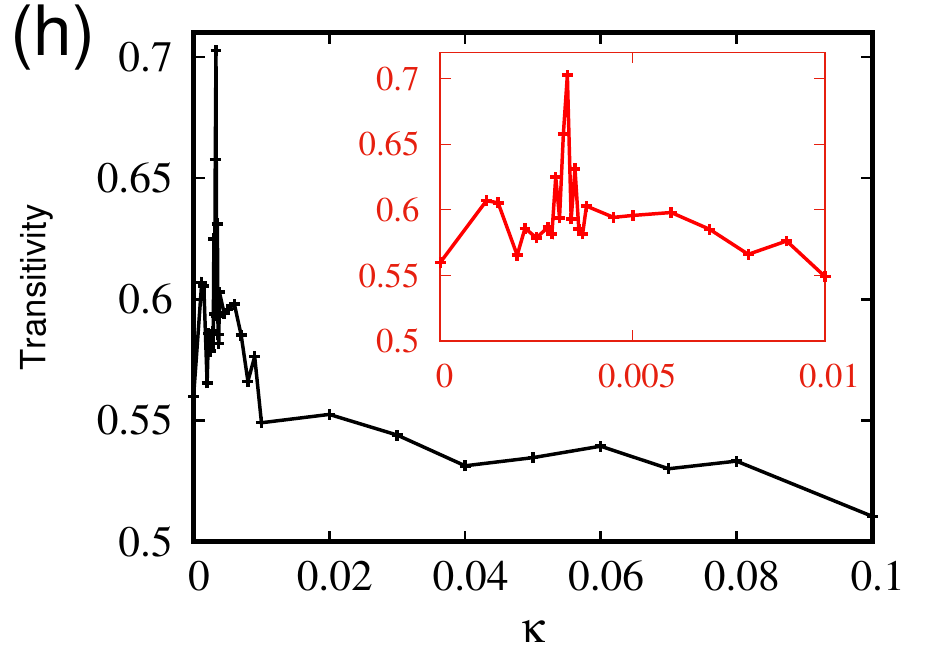}
 \includegraphics[height=3cm, width=5.5cm]{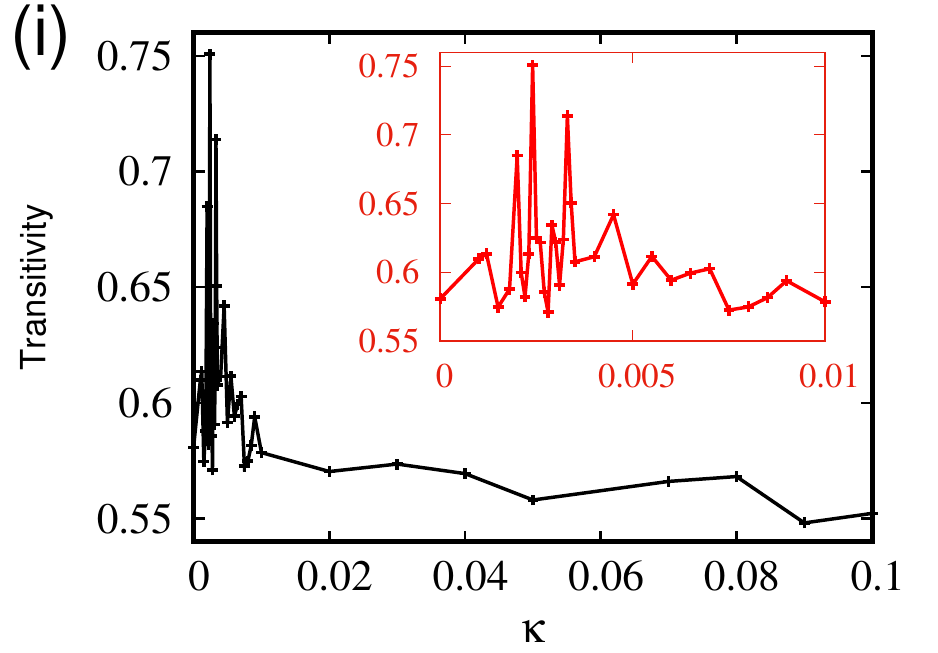}
\vspace{-1ex}
 \caption{Top panel: $\aver{N_{1}}$ versus  $\tau$ 
 for different values of $\kappa$,  with $\vert\alpha\vert^{2}$ 
 equal to (a) $20$, (b) $25$ and (c) $30$.  
 Clustering coefficient versus $\kappa$ (center panel) and transitivity versus $\kappa$ (bottom panel) for (d) and (g) $\vert\alpha\vert^{2} = 20$, (e) and (h) $\vert\alpha\vert^{2} = 25$, and (f) and (i) $\vert\alpha\vert^{2} = 30$.}
 \label{fig:apl_den_cc_kappa}
\end{figure*}

\begin{figure}
 \centering
\includegraphics[height=2.3cm, width=4.24cm,angle=-00]{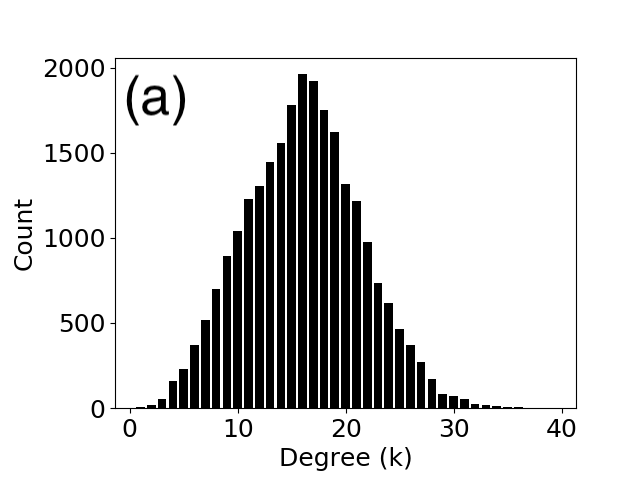}\hspace{0ex}
 \includegraphics[height=2.3cm, width=4.24cm,angle=-00]{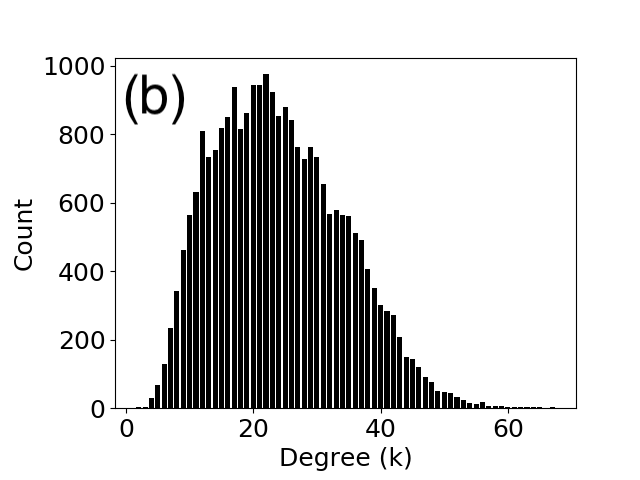}\hspace{0ex}
 \includegraphics[height=2.3cm, width=4.24cm,angle=-00]{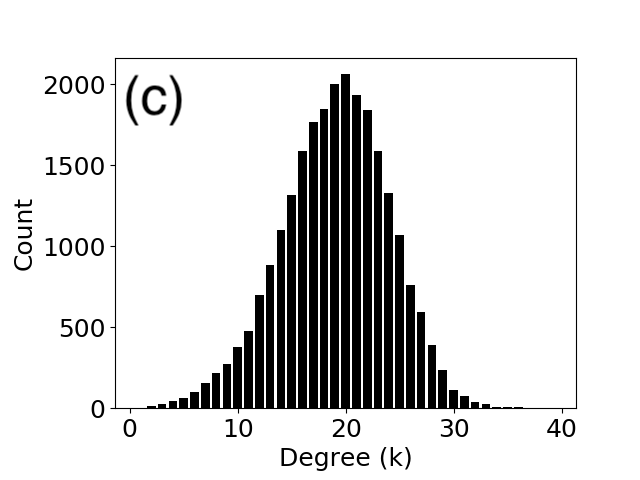}\hspace{0ex}
 \includegraphics[height=2.3cm, width=4.24cm,angle=-00]{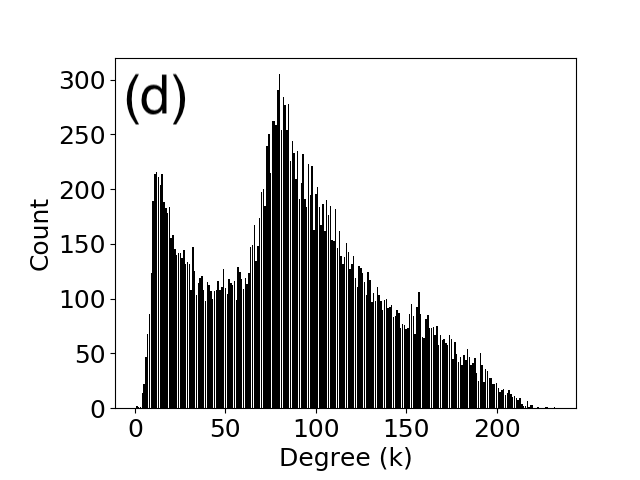}\vspace{-0ex}
 \vspace{-2ex}
 \caption{Degree distribution for $\vert\alpha\vert^{2}=25$ with  $\kappa$ equal to (a) $0$, (b) $0.0033$, (c) $0.01$, and (d) $0.1$.}
 \label{fig:deg_dist}
\end{figure}  
We have carried out extensive numerical investigations of  
the dynamics as a function of $\kappa$ for different values of $|\alpha|^{2}$,  
confirming  in each case  that the special value of 
$\kappa$ identified in the short-time dynamics  retains its distinct nature 
 in the long-time dynamics as well, as reflected in the clustering 
 coefficient  obtained  from $\epsilon$-recurrence  network analysis.  
   The pinching effect in the neighbourhood of the  special value of $\kappa$ is 
   evident  in Figs.  \ref{fig:apl_den_cc_kappa}(a)-(c) depicting 
 the short-time  dynamics for $|\alpha|^{2} = 20, 25$ and $30$, respectively.  
 The corresponding dependence of   CC and transitivity on  $\kappa$ for fixed $\epsilon_{c}$ is shown in Figs.  \ref{fig:apl_den_cc_kappa}(d)-(i). (The precise value of $\epsilon_{c}$ obviously depends on the value of $\kappa$). The  spectacular feature that once again emerges is that at precisely these values of $\kappa$ ($= 0.0066, 0.0033$ and $ 0.0024$  
 respectively for  $\vert\alpha\vert^{2} = 20$, $25$ and $30$) obtained from the appearance of the bifurcation cascade,  the 
 corresponding clustering coefficient is at its  maximum (Figs.  \ref{fig:apl_den_cc_kappa}(d)-(f)). 
 The insets in Figs.  \ref{fig:apl_den_cc_kappa}(d)-(i)  highlight the peaks in both CC and transitivity at the special values of $\kappa$ for different values of  $\vert\alpha\vert^{2}$.


An understanding of the bifurcation cascade and the importance of the special value of $\kappa$ can be obtained by first considering the classical Duffing equation   which describes  an externally driven beam  supported in a frame, on which an axial compressive force $P$ and  a horizontal harmonic excitation force $f\, \cos(\omega\,E\,t)$  act.
(See Ref. \cite{john_argyris}, p. 738, Fig. 10.5.1.)
 Sensitivity to parameter values is manifested via a series of bifurcations.

We draw attention to the `imploding sombrero'  and the oscillatory behaviour ({\it ibid}., p. 745, Fig. 10.5.6) that this cubic nonlinearity produces. In our system the beam with its two arms and the connecting cross-bar is analogous to the three-level atom.  The  two external forces (one on each arm and the cross-bar) are analogous to the two fields that interact between two levels of the three-level atom. However the nature of the square-root nonlinearity in the quantum system renders the situation more complex and movement from the imploding sombrero (the pinch-effect) to oscillations is through a significant increase in fluctuations at the special value of $\kappa$. Thus this value separates two types of behaviour, and is therefore important in the quantum dynamics of the system considered. The reconstructed long-term dynamics also highlights this special value as CC and transitivity maximise at that value.
 

\section{\label{sec:conclusion} Concluding remarks}
We have applied the ideas of $\epsilon$-recurrence network theory to analyse the long-time dynamics of an observable in a generic model of a tripartite quantum system: a three-level atom interacting with two initially coherent radiation fields.
In its short-time dynamics, this observable (the mean photon number 
$\aver{N_{1}(t)}$ of one of the fields) exhibits a bifurcation cascade  as a function 
of the parameter 
$\kappa$  that characterises an intensity-dependent  atom-field coupling or interaction. 
This feature enables the identification of a special value of $\kappa$. 
 In oder to study  the long-time dynamics of the same observable,  
we have carried out a detailed analysis of the time series of 
$\aver{N_{1}}$ by creating an appropriate $\epsilon$-recurrence network from the series.  We then find  that this specific   value of the bifurcation parameter 
$\kappa$ 
also plays a special role in the long-time dynamics.  
The inference is corroborated by the behaviour 
 of the clustering coefficient obtained from the network. It would be interesting to 
 investigate  the counterparts 
of the foregoing features in the dynamics of other systems, including 
 multipartite ones. 

\acknowledgments  
SL thanks G. Ambika  for  a discussion of $\epsilon$-recurrence 
networks in classical  chaotic systems.


\bibliography{reference}

\end{document}